\documentclass[sigconf,edbt]{acmart-edbt2024}

\def\BibTeX{{\rm B\kern-.05em{\sc i\kern-.025em b}\kern-.08em
    T\kern-.1667em\lower.7ex\hbox{E}\kern-.125emX}}
    
\usepackage{xcolor}
\usepackage{enumitem}
\usepackage{hyperref}
\usepackage{subfig}
\usepackage{graphicx}
\usepackage{makecell} 
\usepackage{booktabs} 
\usepackage{caption} 
\usepackage{fancyvrb} 
\usepackage{subfiles} 
\usepackage{xspace}

\newcommand{\rev}[1]{\textcolor{black}{{#1}}}

\newcommand{\system}{{\sc Galois}\xspace} 
\newcommand{\stitle}[1]{\vspace{1ex}\noindent{\bf #1}}
\newcommand{\ititle}[1]{\vspace{1ex}\noindent{\it #1}}

\usepackage{pgfplots}
\usepgfplotslibrary{groupplots}

\setcopyright{rightsretained}

\acmDOI{}

\acmISBN{978-3-89318-094-3}

\acmConference[EDBT 2024]{27th International Conference on Extending Database Technology (EDBT)}{25th March-28th March, 2024}{Paestum, Italy}
\acmYear{2024}

\begin{document}
\title{Querying Large Language Models with SQL}

\author{Mohammed Saeed}
\email{mohammed.saeed@eurecom.fr}
\affiliation{%
  \institution{EURECOM}
    \country{France}
}

\author{Nicola De Cao}
\email{ndecao@google.com}
\affiliation{%
  \institution{Google AI}
  \country{UK}}

\author{Paolo Papotti}
\email{papotti@eurecom.fr}
\affiliation{%
  \institution{EURECOM}
  \country{France}}

\begin{abstract}
In many use-cases, information is stored in text but not available in structured data. However, extracting data from natural language (NL) text to precisely fit a schema, and thus enable querying, is a challenging task. With the rise of pre-trained Large Language Models (LLMs), there is now an effective solution to store and use information extracted from massive corpora of text documents. Thus, we envision the use of SQL queries to cover a broad range of data that is not captured by traditional databases (DBs) by tapping the information in LLMs. 
This ability would enable the hybrid querying of both LLMs and DBs with the SQL interface, which is more expressive and precise than NL prompts. 
To \rev{show the potential} of this vision, we present \rev{one possible direction to ground it with} a traditional DB architecture using physical operators for querying the underlying LLM. One \rev{promising} idea is to execute some operators of the query plan with prompts that retrieve data from the LLM. For a large class of SQL queries, querying LLMs returns well structured relations, with encouraging qualitative results.  
We pinpoint several research challenges that must be addressed to build a DBMS that jointly exploits LLMs and DBs. While some challenges call for new contributions from the NLP field, others offer novel research avenues for the DB community.
\end{abstract}

\maketitle

\section{Introduction}
\textit{Declarative querying} is one of the main features behind the popularity of database systems. 
However, SQL can be executed only on structured datasets with a well defined schema, leaving out of immediate reach information expressed as unstructured text.

Several technologies have been deployed to extract structured data from unstructured text and to model such data in relations or triples
~\cite{ZhangRCSWW17,systemt}. While these methods have been studied for more than 20 years, creating well-formed structured data from text is still time consuming and error prone. 
{Existing tools require engineers to \rev{manually} prepare extraction pipelines, which are typically
static and can only extract fixed \rev{pairs} of attributes~\cite{ZhangRCSWW17}}. \rev{Creating such pipelines is expensive, as training examples must be defined for every relation to extract.} Indeed, the precise extraction of typed data in a 
tuple format (\rev{n-ary relations)} is still an unsolved task~\rev{\cite{3524284,peng2017cross}}.

\begin{figure}[t]
    \centering
    \includegraphics[width=0.48\textwidth]{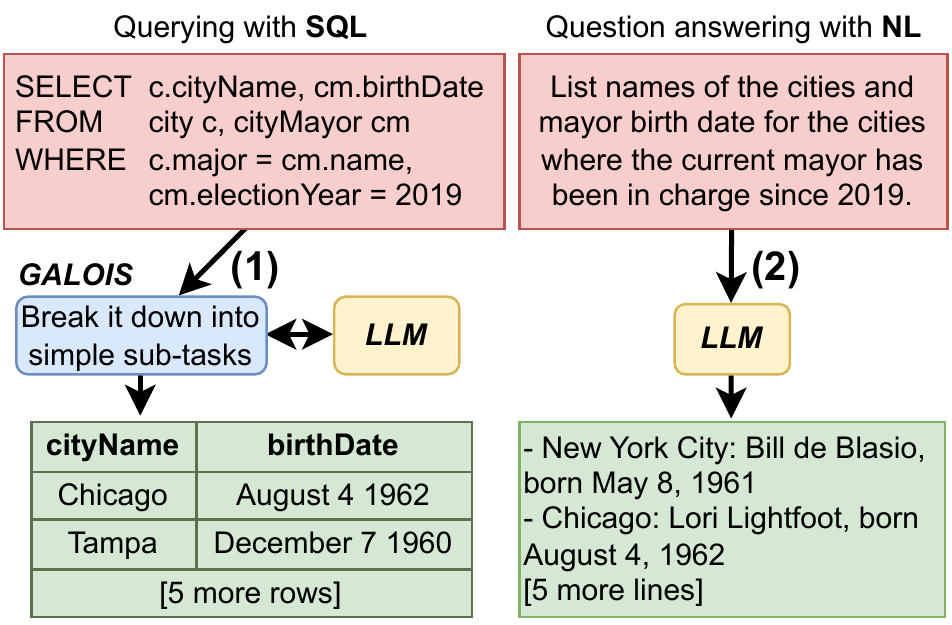}
    \caption{Querying a pre-trained LLM with SQL is different from question answering (QA). We assume a \rev{user} SQL query as input. \system executes the query, and obtains relations, by retrieving data from a LLM (1). The corresponding QA task consumes and produces natural language text (2). 
    }
    \label{fig:example}
\end{figure}

\stitle{Querying vs QA.} While declarative querying of text is a big challenge, there has recently been incredible progress in \textit{question answering} (QA) over text~\cite{rogers2023qa}. In this setting, a question in natural language (NL) is answered by gathering information from a corpus of text documents. Transformers enable the creation of Large Language Models (LLMs), neural networks that are used in a wide variety of NL processing tasks. LLMs, such as those in the GPT family~\cite{radford2018improving,radford2019language,brown2020language}, have been trained on large data, such as the entire Web textual content, and can answer complex questions  in a closed-book fashion~\cite{roberts-etal-2020-much} (example (2) in Figure~\ref{fig:example}). Question answering is reaching new state of the art performance with the release of new LLMs, but it is still not possible to query, in a SQL-like declarative fashion, such models. While it has been shown that such models store high quality factual information~\cite{petroni-2019-language,6b493230}, they are not trained to answer complex SQL queries and may fail short with such input. 

\stitle{SQL for LLMs.} 
We envision querying pre-trained LLMs with SQL scripts. 
As depicted in example (1) in Figure~\ref{fig:example}, the pre-trained LLM 
can act as the data storage containing the information to answer the query. 
We argue that a solution should preserve the main characteristics of SQL when executed over this new source of data: 
(i) queries are written in arbitrary SQL over a user defined relational schema, enabling a precision and a complexity in contrast with the limitation of NL prompts;
(ii) answers are \textit{correct} and \textit{complete} w.r.t. the information stored in the LLM. This last point requires the correct execution of the queries and does not assume that LLMs always return perfect information. In contrast to generation of images or fiction, where small errors are unnoticeable by users in most cases, any error in data can be critical for the target application. While LLMs still make factual mistakes, this work shows that it is already possible to collect tuples from them with promising results. With the ongoing efforts in LLMs, with new training architectures and increasing amount of text used as input, there is evidence that their factuality and coverage is quickly improving~\cite{ElazarKRRHSG21,arxiv.2211.08412}.

\begin{figure}[t]
    \centering
    \includegraphics[width=0.4\textwidth]{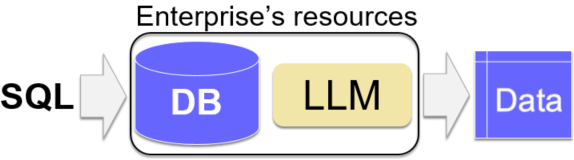}
    \caption{With the increasing number of enterprise LLMs trained with proprietary data, hybrid SQL querying enables to extract structured data from heterogeneous sources. The DB models the relational data, while the LLM exposes the data from unstructured sources. \rev{This paradigm can query data in text without human preprocessing.}
    }
    \label{fig:hybrid}
\end{figure}

\stitle{Applications.}
We envision the execution of SQL queries to obtain relations from the information stored in LLMs. 
Given the increasing adoption of proprietary LLMs by companies, 
domain-specific textual information is getting stored in such models~\cite{wu2023bloomberggpt,DatabricksBlog2023DollyV2}. This is a promising solution for several applications. In any domain, 
data is scattered across different modalities such as email, text, and PDF files. 
Querying their representation in a LLM enables the retrieval of information that is out of reach by accessing only the DB, as depicted in Figure~\ref{fig:hybrid}. 
We envision the ability to query 
data beyond what is already modeled in a structured schema, for example by designing a polystore system on 
heterogeneous storage engines~\cite{AgrawalCCEIKKLM18,duggan2015bigdawg}. 
\rev{An example of jointly querying with one user-provided script a DB and a LLM is the following} 
\begin{verbatim} 
q: SELECT c.GDP, AVG(e.salary)
   FROM LLM.country c, DB.Employees e
   WHERE c.code = e.countryCode
   GROUP BY e.countryCode
\end{verbatim} 
where \textit{c} iterates over the tuples in the LLM and \textit{e} over the tuples in the DB.
With hybrid querying, the data from the LLM can be used as a source in metadata inference~\cite{DengSL0020}, data integration~\cite{golshan2017data}, augmentation~\cite{zhao2022leva}, imputation\cite{MeiSFYFL21}, and cleaning~\cite{wrangle22}. 
This paper does not claim to detail a concrete solution to all these applications, but rather to show a possible path to combine traditional DBMSs and LLMs in novel hybrid query execution plans~\cite{KaoudiQ22}.


\stitle{\rev{Which Architecture?}} 
\rev{Being able to SQL query LLMs is appealing, but it is not clear on which architecture a solution should pivot on. Looking at the architectures for LLMs and DBMSs, there are different paths to explore. One is \textit{LLM-first}, where external information (including structured data) is accessed by the LLM~\cite{pmlr-v162-borgeaud22a,peng2023check}. While this approach is gaining visibility, the limited context size in LLMs does not allow yet to execute queries that require a large number of tuples as input, such as aggregate queries over tables with thousands of rows.
The alternative path is \textit{DB-first}, where LLMs are used as a component in a traditional DB query processing architecture, which is what we envision in this work.} 

\stitle{Contributions.} In this paper, we focus on how to query pre-trained LLMs (in isolation) and preliminary empirical evidence of its potential. \rev{We present one possible way to implement a DB-first architecture.} {The core idea is that the query plan is a natural decomposition of the (possibly complex) process to obtain the result, in analogy with the recent approaches in NLP showing that breaking a complex task in a chain of thoughts is key to get the best results~\cite{CoT22,khattab2022}}. 
To bridge the gap between a logical query plan and its execution on a LLM, we suggest new physical operators for such plan. 

Our \rev{main ideas are} summarized in the following points:
\begin{itemize}
    \item We introduce the problem of querying with SQL existing pre-trained LLMs. 
    To ground our vision, we built \system\footnote{\'Evariste Galois (rhymes with French word \textit{voil\`a}) was a 19$^{th}$ century mathematician.}, a \rev{DB-first} prototype that executes SPJA queries under assumptions that enable a large class of applications (code available at~\url{https://gitlab.eurecom.fr/saeedm1/galois}).
    
    \item {The logical query plan breaks down the complex task into simpler steps that can be handled effectively by the LLM}. Physical operators in the query plan are implemented as textual prompts for LLMs. \rev{Such prompts are automatically generated from the information in the input schema and the logical operators.}
    
    \item We show that the results obtained by \system on 46 queries on top of popular LLMs are (1) comparable to those obtained by 
    executing the same queries on 
    DBMS and (2) better 
    than those obtained by manually rewriting the queries (and parsing the results) in NL for QA over the same LLM.
\end{itemize}

\stitle{Outline.}
Section~\ref{sec:background} covers recent progress in NLP 
and compares \system to prior work. Section~\ref{sec:design} discusses the challenges in querying LLMs. Section~\ref{sec:overview} describes the architecture of our \rev{DB-first} 
prototype. Section~\ref{sec:experiments} reports preliminary experimental results from datasets in the Spider corpus. Section~\ref{sec:directions} discusses open problems and \rev{research directions}. Section~\ref{sec:Conclusion} concludes the paper.

\section{Background}
\label{sec:background}
Our vision is inspired by recent advances in the domain of natural language processing (NLP). 
Progress in this field has been driven by two major concepts: the Transformer neural network architecture and the application of transfer learning~\cite{devlin2018bert}. 
One of the transformer's benefits is its suitability for parallelization w.r.t. previous approaches, which has enabled the creation of massive LLMs \cite{brown2020language}. These models are pre-trained on tasks, such as predicting the next word in a sentence, for which large amounts of data are easily accessible. Although pre-training is costly, the models can then be adapted to 
new tasks.
Traditionally, ``fine-tuning'' with annotated examples for a target task has been the main way of customizing pre-trained LLMs. However, the latest generation of pre-trained models has opened up new possibilities. Models of sufficient size complete new tasks without any additional training, simply by being given NL descriptions of the task ("instruction tuning``).
Precision is improved by incorporating a limited number of examples (e.g., five to ten) that pair the input for the task with its solution ("few-shot learning"). 
An example of a 
prompt for GPT-3 is a question in natural language (``what is the capital of USA?") or a request 
(``The EU state capitals are:").

Our effort is different from the problem of semantic parsing, i.e., the task of translating NL questions into 
SQL~\cite{qatch,spider,Katsogiannis-Meimarakis21}.
Our goal is also different from querying an existing relational database to answer a NL question~\cite{HerzigNMPE20}.
We are interested in retrieving data from the LLM with SQL queries, with the traditional semantics and with the output expressed in the relational model, as if the query were executed on a DBMS. While some of these facts can be retrieved with QA, (i) the SQL query must be rewritten as an equivalent question in NL, which is not practical for complex scripts, (ii) the textual result must be parsed into a relation, (iii) 
current LLMs in some cases fail in answering complex queries expressed as NL. 
{Indeed, QA systems are optimized for answering questions with a text, while SQL queries return results in the form of tuples, possibly with complex operations to combine intermediate values, such as aggregates, where LLMs fail short~\cite{RibeiroWGS20}.
To overcome some of these limits, it has recently been shown that a series of intermediate reasoning steps (``chain of thought'' and question decomposition~\cite{tacl_a_00309}}) improve LLMs' ability in complex tasks~\cite{CoT22}.

Our work is also different from the recent proposal for Neural DBs~\cite{ThorneYSS0L21}, where textual facts are encoded with a transformer and queries are posed with short
NL sentences. We do not assume facts as input and we focus on traditional SQL scripts executed on LLMs.

Using an LLM query as a component in SQL query answering has analogies with other DBMS extensions in the literature, such as involving crowed workers to answer open world questions~\cite{crowddb}.

\section{Design Considerations}
\label{sec:design}
Our goal is to execute SQL query over the data stored into LLMs.
When we look at these models from a DB perspective, they (i) have extensive coverage of facts from massive textual sources; 
(ii) have perfect availability; 
(iii) directly query a very compressed version of the data, as facts are stored effectively in the parameters of the model: {the CommonCrawl+ text corpus takes 45TB, while GPT-3 only 350GB}.
However, LLMs have their shortcomings, as we discuss next, including poor data manipulation skills, e.g., they fail with numerical comparisons. Conversely, traditional query operators are great at processing data with rich operators, such as joins and aggregates, but only within the data available in the given relation. 

The combination of LLMs and traditional DBMSs shows the potential for a hybrid system that can jointly query existing relational tables and facts in the LLM. However, it is crucial to consider the limitations and challenges in querying LLMs. We now delve into three key issues that have impacted the design of \system.

\stitle{1. Tuples and Keys.}
As far as we know, LLMs do not have a concept of schema or tuple, but they model existing relationships between entities (``Rome is located in Italy'') or between entities and their properties (``Rome has 3M residents''). 
However, a query asking for city names may assume that a name identifies a city, which is not the case in reality, e.g., there is a Rome city in Georgia, USA. 

In some cases, key attributes exist in the real world. 
For example, LLMs contain keys such as IATA codes for airports. , e.g., `JFK`. 
However, in general, we do not have a universal global key for several entities, such as cities, and the default semantics for the LLM is to pick the most popular interpretation, with popularity defined by occurrences of terms in the original pre-training data.

In general, this problem can be solved with keys defined with multiple attributes, i.e., the \textit{context} in NLP terminology. For example a composite key defined over (name, state, country)
enables us to distinguish the Rome in Italy from the one in Georgia.
In our initial prototype, we assume that every relation involved in the query has a key and that the key can be expressed with one attribute, e.g., its name. This constraint can be relaxed by handling composite keys.

\stitle{2. Schema Ambiguity.}
A major challenge in language is ambiguity. 
Similarly to the issue with entities, several words, including attribute labels, can have multiple meanings. These alternatives are represented differently in the parameters of LLMs. In our setting, 
a given attribute label in the query can be mapped to multiple ``real world'' attributes in the LLM, e.g., \textit{size} for a city can refer to population or urban area~\cite{VeltriSB0P22}. 

In this initial effort, we assume that meaningful labels for attributes and relations are used in the queries. This allows the system to obtain prompts of good quality automatically. 

%

\stitle{3. Factual Knowledge in LLMs.}
LLMs do \textit{not know what they know}. This is an intrinsic challenge in the transformer architecture and the decoder, specifically. The decoder returns the next token in a stream. Such token may be based on either reliable acquired knowledge, or it may be a guess. For this reason, a query result obtained LLMs is not 100\% reliable and cannot be immediately verified as LLMs do not expose their sources with the results. 
However, with \system, we experimentally demonstrate that it is possible to extract factual information from LLMs to answer SQL queries. Moreover, new models keep increasing the factuality of their answers\footnote{For example, ``GPT-4 scores 40\% higher than our latest GPT-3.5 on our  factuality evaluations'' - \url{https://openai.com/research/gpt-4} - published on March 15$^{th}$ 2023}. In this work, we do not tackle the general problem of separating the \textit{knowledge about language and reasoning} from \textit{factual knowledge}, which is an ongoing NLP research topic as we discuss in Section~\ref{sec:directions}.

\section{Overview}
\label{sec:overview}
The high-level architecture of \system is presented in 
Figure~\ref{fig:example}. {We assume that the schema (but no instances) is provided together with the query}. The system processes SQL queries over data stored in a pre-trained LLM. This design enables developers to implement their applications in a conventional manner, as the complexities of using an LLM are encapsulated within \system. 



\begin{figure}
    \centering
    \includegraphics[scale=0.28]{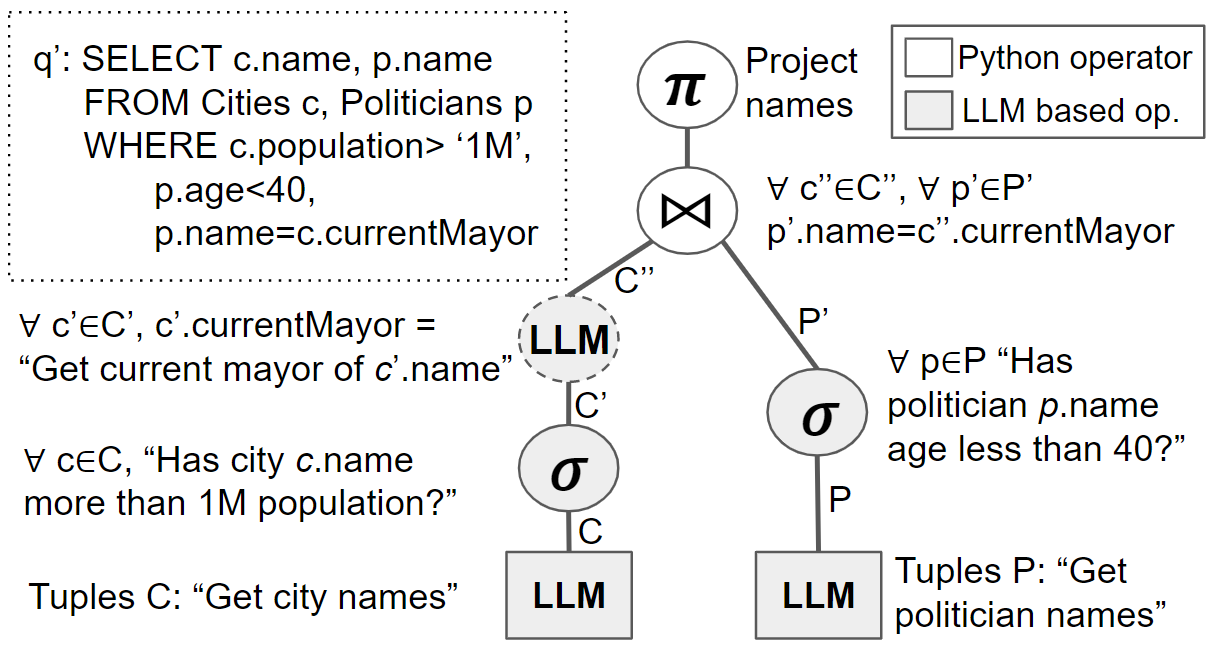}
    \caption{Logical plan for query q'. 
    Base relations are accessed by retrieving sets of tuples (\textit{C}, \textit{P}) with one key attribute (\textit{name}) from the LLM. Other LLM operators consume and produce tuple sets, retrieving for every tuple the required attributes, if not in the tuple yet. 
    The last two operators do not involve the LLM.}
    \label{fig:queryPlan}
\end{figure}

\stitle{Operators.} The core intuition of our approach is to use LLMs to implement a set of specialized physical operators in a traditional query plan, as demonstrated in Figure~\ref{fig:queryPlan}. As tuples are not directly available, we implement the access to the base relations (leaf nodes) with the retrieval of the key attribute values. We then retrieve the other attributes as we go across the plan. For example, if the selection operator is defined on attribute $A$ different from the key, the corresponding implementation is a prompt that filters every key attribute based on the selection condition, e.g., "Has city \textit{c.name} more than 1M population?", where \textit{c.name} iterates over the set of key values. If a join or a projection involve an attribute that has not been collected for the tuple, this is retrieved with a special node injected right before the operation. For example, if a join involves an attribute ``currentMayor'', the corresponding attribute values are retrieved with a prompt that collects it for every key, such as ``Get the current mayor of \textit{c'.name}". Once the tuples are completed, regular operators, implemented in Python in our prototype, are executed on those, e.g., joins and aggregates.

{On one hand, the query plan acts as a chain of thought decomposition of the original task, i.e., the plan spells outs intermediate steps. On the other hand, the operators that manipulate data fill up the limitations of LLMs, e.g., in computing average values or comparing quantities~\cite{quantitative22}. Together, these two features make the LLM able to execute complex queries.}

\stitle{Prompts.} Figure~\ref{fig:queryPlan} shows how prompts, suitable for execution on LLMs, implement logical operators in \system. 
A prompt is obtained for each operator by combining a set of operator-specific prompt templates with the labels/selection conditions in the given SQL query. In the example query $q'$, politicians ({\small \fontfamily{phv}\selectfont Politicians p}) are filtered according to their age ({\small \fontfamily{phv}\selectfont p.age<40}); this corresponds to retrieving a set of tuples (P) with one key attribute (name), which is then followed by a prompt that for each politician checks in the LLM its age. For example, we instantiate the template ``Has \textit{relationName} 
\textit{keyName} \textit{attributeName} \textit{operator} \textit{value}?'', with `politician' , `B. Obama', `age', `less than', `40', respectively. 
\rev{Templates are a simple solution for prompting. However, DB-first approaches such as \system can directly exploit results in prompt engineering~\cite{10.1145/3560815}, as those could be plugged in step (2) of the workflow described next.}

\stitle{Workflow.} The main operations in \system's query processing are: 

\begin{enumerate}
    \item Obtain a logical query plan for a query $q$ and the source schema. We assume the label of the key attribute is given. 
    \item Access the LLM to retrieve the tuples composed of the key attribute and to gather more attributes in case of selection, join and projection. Each operation is done with a prompt template filled up with the labels and conditions at hand. 
    \item Convert the string of answers from the LLM to a set of CELL values in the attribute.
    \item Use traditional algorithms for any operator involving attributes that have already been retrieved.
\end{enumerate}

Two critical steps enable the practical use of \system.
First, as relations can be large, we iterate with the a prompt until we stop getting new results. For example, we  ask for city names, collect the answer in a set, and keep asking for more names with another prompt (``Return more results''). The termination condition could be replaced by a user-specified threshold. 
A second issue is the cleaning of the data gathered from the LLM. For example, numerical data can be retrieved in different formats. We normalize every string expressing a numerical value (say, 1k) into 
a number (1000). The enforcing of type and domain constraints is a simple but crucial step to 
limit the incorrect output due to model hallucinations. \rev{More LLM-specific methods can be plugged for this cleaning step. For example, the generated output can be critiqued by another model~\cite{shinn2023reflexion,mündler2023selfcontradictory}, or the LLM can be augmented with external information~\cite{surveyHallucination}.
}

\rev{In contrast with the traditional approach based on extracting structured data from text, LLMs and prompts enable the extraction of such data without human annotations. In a DB-first approach, prompts are automatically generated, thanks to the logical plan.}

\section{Experiments}
\label{sec:experiments}
All experiments are executed on popular LLMs for a set of SQL queries for which we have the ground truth according to a database.

\stitle{\rev{Implementation.}}
\rev{\system is written in Python and all LLMs have been executed locally with the exception of ChatGPT, for which we used the API. Query plans are obtained from DuckDB. Code and datasets are available at
~\url{https://gitlab.eurecom.fr/saeedm1/galois}.}
This initial implementation serves as the experimental platform to show the promise of the vision, rather than a full-fledged solution. 

\stitle{Dataset.}
Spider is a Text2SQL dataset with 200 databases, each with a set of SQL queries~\cite{spider}. For each query, it provides its paraphrases as a NL question.
We focus on a subset of \textit{46 queries for which \rev{we expect to} obtain answers from an existing LLM}. \rev{More precisely,} we leave out queries that are specific to the \rev{relational dataset provided by} Spider (e.g., ``How many heads of the departments are older than 56?'') and use in our evaluation only queries \rev{about generic topics}, such as world geography and airports (``What are the names of the countries that became independent after 1950?'').
If there are multiple paraphrases for a question, we pick the first one.

\begin{figure}[t!]
    \centering
\noindent\fbox{%
    \parbox{0.95\columnwidth}{%
    I am a highly intelligent question answering bot. If you ask me a question that is rooted in truth, I will give you the short answer. If you ask me a question that is nonsense, trickery, or has no clear answer, I will respond with "Unknown". If the answer is numerical, I will return the number only.
    
Q: What is human life expectancy in the United States?

A: 78.
 
Q: Who was president of the United States in 1955?

A: Dwight D. Eisenhower.
 

 
 Q: What is the capital of France?

 A: Paris.

Q: What is a continent starting with letter O?

A: Oceania.
 
Q: Where were the 1992 Olympics held?

A: Barcelona.
 
Q: How many squigs are in a bonk?

A: Unknown
}%
}
    \caption{
    Few shot examples for the 
     GPT-3's prompt.}
    \label{fig:prompt}
\end{figure}

\stitle{Setup.}
We test four LLMs. \textit{Flan-T5-large} (\textbf{Flan}): T5 fine-tuned on 
datasets described via instructions (783M parameters). \textit{TK-instruct-large} (\textbf{TK}): T5 with instructions and few-shot with positive and negative examples (783M parameters). \textit{InstructGPT-3} (\textbf{GPT-3}): fine-tuned GPT-3 using instructions from humans~\citep{Ouyang2022TrainingLM} (175B parameters). \textit{GPT-3.5-turbo} (\textbf{ChatGPT}): chat model in the OpenAI API (175B parameters). {We construct prompts appropriately for each model,}
we report the one for {GPT-3} in Figure~\ref{fig:prompt}, {showing the instruction in the prompt followed by some few-shot examples.}

For a given LLM $M$ and a SQL query $q$ with its Spider relation $D$ and the corresponding NL question $t$, we collect \rev{four} results: 
(a) relation $R_M$ from \system executing $q$ over $M$, 
(b) relation $R_D$ by executing $q$ over $D$,
(c) text $T_M$ by asking $t$ over $M$,
\rev{(d) text $T_M^C$ by asking $t$ over $M$ using a chain-of-thought prompt.}
\rev{The last result (d) explores the middle ground between standard questions answering (c) and \system (a). For NL question $t$, an engineered prompt contains a complete example of a manually crafted chain-of-thought (CoT), similar to the logical plan execution for the query, followed by $t$ and instructions to reason step by step.
In this method, the CoT example in the prompt is fixed as how to derive a decomposition automatically from $t$ is an open problem.}
Only (b) uses the relations from Spider, (a), (c) and \rev{(d)} get the data from the LLM.

\stitle{Evaluation.} 
We analyze the results across two dimensions. 

\ititle{1) Cardinality.} First, we measure to which extent \system returns correct results in terms of number of tuples. As NL questions always return text paragraphs, we cannot include their results in this analysis. 
For \system, all output relations have the expected schema, this is obtained by construction from the execution of the query plan, i.e., every $R_M$ has the same schema as every $R_D$.
However, in terms of number of tuples there are differences.  
We compute the ratio of the sizes as $f=\frac{|2*R_D|}{|R_D+\rev{R_M|}}$, 
\rev{where the interval for $f$ is [0,2] and best result occurs when $R_D==R_M$ ($f$=1).}
\rev{Consider expected Relation $R_D$ with size (3,2), i.e., 3 tuples and 2 columns. Assume \system produced $R_M$ = (1,2). 
In this case, f = |2*3| / (3+1) = 6/4 = 1.5.}

\setlength{\tabcolsep}{3.86pt}
\begin{table}[t]
\centering
\begin{tabular}{lcccc}
\toprule 
  & \textbf{Flan} & \textbf{TK} & \textbf{GPT-3} & \textbf{ChatGPT} \\

\midrule 
 Difference as \% of $R_D$ size  
 & -47.4 & -43.7  & +1.0   & -19.5   \\
\bottomrule
\end{tabular}
\caption{Average difference in the cardinality of \system's output relations \rev{($R_M$)} w.r.t. the ground truth results $|R_D|$ for the 46 Spider queries. 
Closer to 0 is better.}
\label{table:cardinality}
\end{table}

In Table~\ref{table:cardinality}, we report the difference as percentage (averaged over all queries with non-empty results) with the formula 1-$f$. 


Results show that smaller models do worse and miss lots of result rows, up to 47.4\% w.r.t. the size of results from the
SQL execution $R_D$. For GPT models, almost all queries return a number of tuples close to $R_D$. Most differences are explainable with errors in the results of the prompts across the query pipeline, as we discuss 
next.

\begin{table}[t]
\centering
\begin{tabular}{lcccc}
\toprule 
& \textbf{All} & \textbf{Selections} & \textbf{Aggregates} & \textbf{Joins} \\
& \textbf{} & \textbf{only} & \textbf{} & \textbf{} \\
\midrule 
$R_M$ (SQL Queries)  &  50 & 80 & 29  & 0   \\
$T_M$ (NL Questions) &  44  & 71  & 20  & 8     \\
\rev{$T_M^{C}$ (NL Quest.+CoT)}  &  41 & 71 & 13  & 0   \\
\bottomrule
\end{tabular}
\caption{\rev{Cell value matches (\%) between the result returned by a method and the same query executed on the ground truth data ($R_D$) for the 46 {Spider} queries.} Averaged results for \textbf{ChatGPT}.}
\label{table:results}
\end{table}

\ititle{2) Content}. Second, we measure the quality of the 
results  
by comparing the content of each cell value after manually mapping tuples between $R_D$ on one side (ground truth) and (\rev{$R_M$, $T_M$, $T_M^C$}) on the other. 
As \rev{$T_M$, $T_M^C$ contain} NL text, we manually postprocess them to extract the values as records.
\rev{In our manual mapping, we split comma-separated values, remove repeated values and punctuation, and map the resulting tuples to the ground truth records - how to automate this mapping process is an open problem.}
We consider a numerical value in (\rev{$R_M$, $T_M$, $T_M^C$}) as correct if the relative error w.r.t. $R_D$ is less than 5\%.

As this analysis requires to manually verify every result, we conduct it only for one LLM.
Results in Table~\ref{table:results} show that \system executes the queries on ChatGPT with a better average accuracy in the results compared to the same queries expressed as questions in NL. We believe this is a very promising result, as one can think that the results coming from the NL QA task are the upper bound for what the LLM knows. For the easiest subclass of queries, selection-only, the query approach returns correct values in 80\% of the cases. Joins are the most problematic, as we observe failure in the join step due to different formats of the same text, e.g., an attempt to join the country code ``IT" with ``ITA'' for entity Italy. 
\rev{This challenging subclass of queries clearly requires more work to drastically increase the homogeneity of the intermediate results. The results also show that well-engineered chain-of-thought NL prompts (\rev{$T_M^C$}) do not lead to better results than 
\system, confirming the quality of the chain of prompts obtained automatically.}

\vspace{1ex}
As we do not control the infrastructure of OpenAI, we do not report API execution times. On average, {GPT-3} takes $\sim$20 seconds to execute a query ($\sim$110 batched prompts per query). Distributions for these metrics are skewed as they 
depend 
on the result sizes.

\section{Research Directions} 
\label{sec:directions}
\system aims at creating a system that can push the boundaries of declarative query execution over LLMs, while achieving comparable accuracy and performance to queries executed on a traditional DBMS. 
While the current prototype does not yet meet these goals, we discuss the main next steps in this vision, including open research questions and associated challenges.

\stitle{Query optimization.} 
As in a traditional DBMS, optimization can be organized according to the logical and physical plans.

For the \textit{logical} plan,
\rev{an advantage of the DB-first approach is the automatic generation of chain-of-thought prompts. However, the ability to combine in one query plan operators over traditional storage and LLMs is a vision that go beyond the scope of chain-of-thought. For example, an optimizer may be able to recognize when the execution of the (more expensive) LLM is needed at runtime.
We also} need optimization heuristics to obtain equivalent logical plans that reduce the number of prompt executions (which can be large) over the LLM. In the example in Figure~\ref{fig:queryPlan}, pushing down the selection over city population to the data access call (leaf) requires to combine the prompts, e.g., ``get names of cities with $>$ 1M population''. This simple change removes the prompt executions for filtering the list of all cities. However, the optimization decision is not trivial as combining too many prompts lead to complex questions that have lower accuracy than simple ones. 

For the \textit{physical} plan, interesting problems arise around the textual prompts. Research questions include how to generate them automatically given only the attribute labels, especially when those are ambiguous or cryptic. The rule of thumb is that the more precise the prompt, the better will be the accuracy of its results. One direction is to make use of data samples, when available. Giving examples of the desired output would guide the LLM to the right format and encoding, which is an issue in our current implementation. Another approach is to optimize the prompt for the retrieval task, with some fine tuning or by exploiting pre-defined embeddings for the desired attribute types~\cite{zhong2021factual}. \rev{For fine tuning, 
reinforcement learning from human feedback (RLHF) can be explored to better stir the generation towards the factual values that are needed for query processing~\cite{bai2022training}.
For the second approach, given a library of type embeddings such as ``Person'' and ``City'', those can be added to prompts for accurate retrieval~\cite{TEs}.}


\stitle{Knowledge of the Unknown.} 
{To overcome the problem of the results mixing real facts and hallucinations, 
one direction is to verify generated query answers by another model, possibly also build on LLMs. In most cases, verification is easier than generation, e.g., it is easier to verify a proof rather than generate it. Our enforcing of simple domain constraints shows benefit, but there is the need to adapt more general data cleaning techniques~\cite{IlyasC19}.} 

Another direction is retrieval augmented language models, where they design modules that separate the ``language understanding and reasoning" part and ``factual knowledge" part~\cite{nicola_thesis}.  Our prompting is a basic approach to surface facts, but more principled solutions are needed to obtain reliable results~\cite{honovich-etal-2021-q2}.

\stitle{Provenance.}
Retrieval augmented models are also a promising direction to address the fact that LLMs \rev{cannot always precisely} cite the sources, or provenance~\cite{zhang-etal-2020-said}, of their output. 
This is an issue, because it is not possible to judge correctness without the origin of the information. 
\rev{With prompt engineering, LLMs can produce an explanation supporting their output and} 
there are ongoing efforts on linking generated utterances, or values in our case, to sources~\cite{AttQA}. This can also be done through {the generation process or in a post-processing step}~\cite{3330885}.


{\stitle{Schema-less querying.}}
We currently assume the SQL schema as given by the user. An interesting extension is to allow users to query without providing a schema. This removes friction from the user, but raises new challenges. Consider the following two queries.

\begin{verbatim} 
Q1: SELECT c.cityName, cm.birthDate   
    FROM city c, cityMayor cm  
    WHERE c.mayor=cm.name
Q2: SELECT cityName, mayorBirthDate 
    FROM city 
\end{verbatim} 

Both of them collect the names of cities with the birth date of the mayor. 
As the LLMs have no schema, both queries should give the same output when executed, i.e.,  
two SQL queries that are both correct translation of the same NL question should give equivalent results.
How to guarantee this natural property (for DBs) is a challenge that requires to combine the new challenges in the LLM setting with results on SQL query equivalence~\cite{guagliardo2017formal}.

\stitle{\rev{Portability.}}
\rev{As SQL queries are \textit{portable} across DB engines, the same SQL script executes on different LLMs. Differently from the schema-less query case, in this case the query $q$ if fixed, but the LLM changes. If two LLMs are trained on the same data, ideally they should return the same answer for $q$. However, this requirement is hard to achieve because of the non deterministic learning process for LLMs. As a consequence, the same prompt does not give equivalent results across LLMs.}

\stitle{Architecture.} 
\rev{\system is based on a DB-first architecture, where the LLM is plugged in the operators. The alternative LLM-first architecture is also promising, but with different challenges. 
An open question is if LLMs can replace DBMSs by consuming the structured data in a training process or as part of the input context. Research on tabular language models show that we are far from this scenario, mostly due to the limitation to the size of the context~\cite{tacl_survey}, but recent research is addressing this issue~\cite{100k}.
However, despite the progress in LLM research, legal and economic hurdles (e.g., the need for formal guarantees in transactions) would still affect an LLM-first solution and may ultimately limit its impact.}

\stitle{Updates and Cost.}
\rev{In the DB-first approach}, we envision that querying LLMs will be less common than querying traditional DBMSs; LLMs are a source for some use cases, but not a replacement. 
However, training and using LLMs is expensive and energy consuming. 
Given the cost of training, it is not clear how to deal with the continuous creation of new information~\cite{LazaridouKGALTG21}. One short term solution is to update LLMs without retraining~\cite{CaoAT21,meng2022memit}. In the long term, 
cost will be reduced by cheaper training and inference
\footnote{For example, ``Through a series of system-wide optimizations, we’ve achieved 90\% cost reduction for ChatGPT since December" - \url{https://openai.com/blog/introducing-chatgpt-and-whisper-apis} - published on March 1$^{st}$ 2023}. 

\stitle{Coverage and Bias.}
LLMs focus on common and probable cases by design. 
We found that, for some queries, 
missing results are due to their lower popularity, compared to those surfaced by the LLM.
Researchers are focusing more on this {challenge}~\cite{ElazarKRRHSG21,arxiv.2211.08412}.
However, LLMs do well with huge amount of data, which is available only for few languages. While the problem is mitigated with machine translation, i.e., by translating from English to a target language, in terms of factual knowledge there is no clear solution. 
\rev{In general, the impact of training data and how to select high quality sources is getting more attention with proprietary LLMs~\cite{penedo2023refinedweb,gunasekar2023textbooks}.}

LLMs encode biases and stereotypes that are present in observed human language, 
we therefore must be careful when applying these models in real-world applications~\cite{parrots21}. 


\section{Conclusion} 
\label{sec:Conclusion}
This paper presents a vision for the DB community by highlighting the potential of querying Large Language Models with SQL, thereby opening up novel research avenues and opportunities. \rev{We report an example for a DB-first approach that leverages} the power of LLMs in combination with traditional DBMSs to create a hybrid query execution environment. 
As LLM factuality and coverage continues to improve, the integration of these models into database systems will not only enable a wide range of data applications, 
but also inspire new contributions from the NLP field. By showcasing a prototype and envisioning the usage of SQL for querying LLMs, we hope to stimulate further exploration and collaboration between DB and NLP communities, ultimately leading to innovative solutions that unlock previously untapped information from unstructured text data in various domains.

\bibliographystyle{ACM-Reference-Format}
\bibliography{references,ref}
\end{document}